\documentclass[twocolumn,showpacs,preprintnumbers,amsmath,amssymb, prl, reprint]{revtex4-1}
\usepackage{amsmath}
\usepackage{stmaryrd}
\usepackage{txfonts}
\usepackage{amssymb}
\usepackage{mathrsfs}
\usepackage{graphicx}
\usepackage{dcolumn}
\usepackage{bm}
\usepackage{epsfig}
\usepackage{color}
\usepackage{hyperref}

\begin{document}
\preprint{\href{http://dx.doi.org/10.1103/PhysRevLett.112.187203}{S.-Z. Lin, C. D. Batista, C. Reichhardt and A. Saxena , Phys. Rev. Lett. {\bf 112}, 187203 (2014).}}

\title{AC current generation in chiral magnetic insulators and skyrmion motion induced by the spin Seebeck effect}
\author{Shi-Zeng Lin}
\affiliation{Theoretical Division, Los Alamos National Laboratory, Los Alamos, New Mexico
87545, USA}
\author{Cristian D. Batista}
\affiliation{Theoretical Division, Los Alamos National Laboratory, Los Alamos, New Mexico
87545, USA}
\author{Charles Reichhardt}
\affiliation{Theoretical Division, Los Alamos National Laboratory, Los Alamos, New Mexico
87545, USA}
\author{Avadh Saxena}
\affiliation{Theoretical Division, Los Alamos National Laboratory, Los Alamos, New Mexico
87545, USA}

\begin{abstract}
We show that a temperature gradient induces an ac  electric current in multiferroic insulators when the sample is embedded in a circuit. We also show that a thermal gradient can be used to move  magnetic skyrmions in insulating chiral magnets: the induced magnon flow from the hot to the cold region drives the skyrmions in the opposite direction  via a magnonic spin transfer torque. Both results are combined to compute the effect of skyrmion motion on the ac current generation and demonstrate that skyrmions in insulators are a promising route for spin caloritronics applications.

\end{abstract}
  %AC current generation in chiral magnetic insulators and skyrmion motion induced by the spin Seebeck effect
\date{\today}
\maketitle

The possibility of inducing magnon currents with thermal gradients in magnetically ordered insulators is attracting considerable attention for its promise in spin caloritronics.~\cite{Bauer2012} For finite magnetoelectric coupling, the magnon current  creates oscillations in the electric polarization, implying that an oscillating current is generated  if the sample is embedded in an electric circuit.  This simple observation suggests that the  ability to induce spin currents with thermal gradients can also be exploited for other applications. The recent discovery of stable spin textures called skyrmions in magnetic insulators without inversion symmetry (e.g. $\mathrm{Cu_2OSeO_3}$)~\cite{Seki2012,Adams2012} posed a challenge for  moving and manipulating these textures in insulating materials.  Because $\mathrm{Cu_2OSeO_3}$ exhibits a finite magnetoelectric coupling, it is important to study how skyrmions respond to a thermal gradient and the effect of skyrmion motion on the electric polarization.

Skyrmions were predicted to be stable in chiral magnets without inversion symmetry. \cite{Bogdanov89,Bogdanov94,Rosler2006} The triangular skyrmion crystal was first observed experimentally in metallic bulk \cite{Muhlbauer2009} and thin films \cite{Yu2010a,Yu2011}. In a metal, conduction electrons interact with the local magnetic moments and drive the skyrmion via spin transfer torque. \cite{Zang11} The threshold current to make skyrmions mobile is low,  \cite{Jonietz2010,Yu2012,Schulz2012} thus skyrmions are very promising for applications in spintronics.

Dissipation due to the conduction electrons is absent in insulators, where the dominant dissipation mechanism is the weak Gilbert damping of the spin precession. Therefore, skyrmions that emerge in  insulating materials  are attractive for applications that require low energy dissipation. Furthermore, the finite magnetoelectric coupling that is intrinsic to skyrmion textures opens the possibility of manipulating these magnetic structures with external electric fields. \cite{Seki2012,Seki2012b} \footnote{The rotation of skyrmion lattice in the presence of a dc electric field in equilibrium was reported in Ref. \onlinecite{White2012}. The electric field couples to the electric polarization associated with the skyrmion lattice, and the new equilibrium state is the rotated lattice configuration.} In analogy with the thermal control of magnetic domain walls in insulators~\cite{Xiao2010,Hinzke2011,Yan2012,Kovalev2012,Bauer2012,Jiang2013}, we propose to drive skyrmions with  a magnon current induced by a thermal gradient. 

\begin{figure}[b]
\psfig{figure=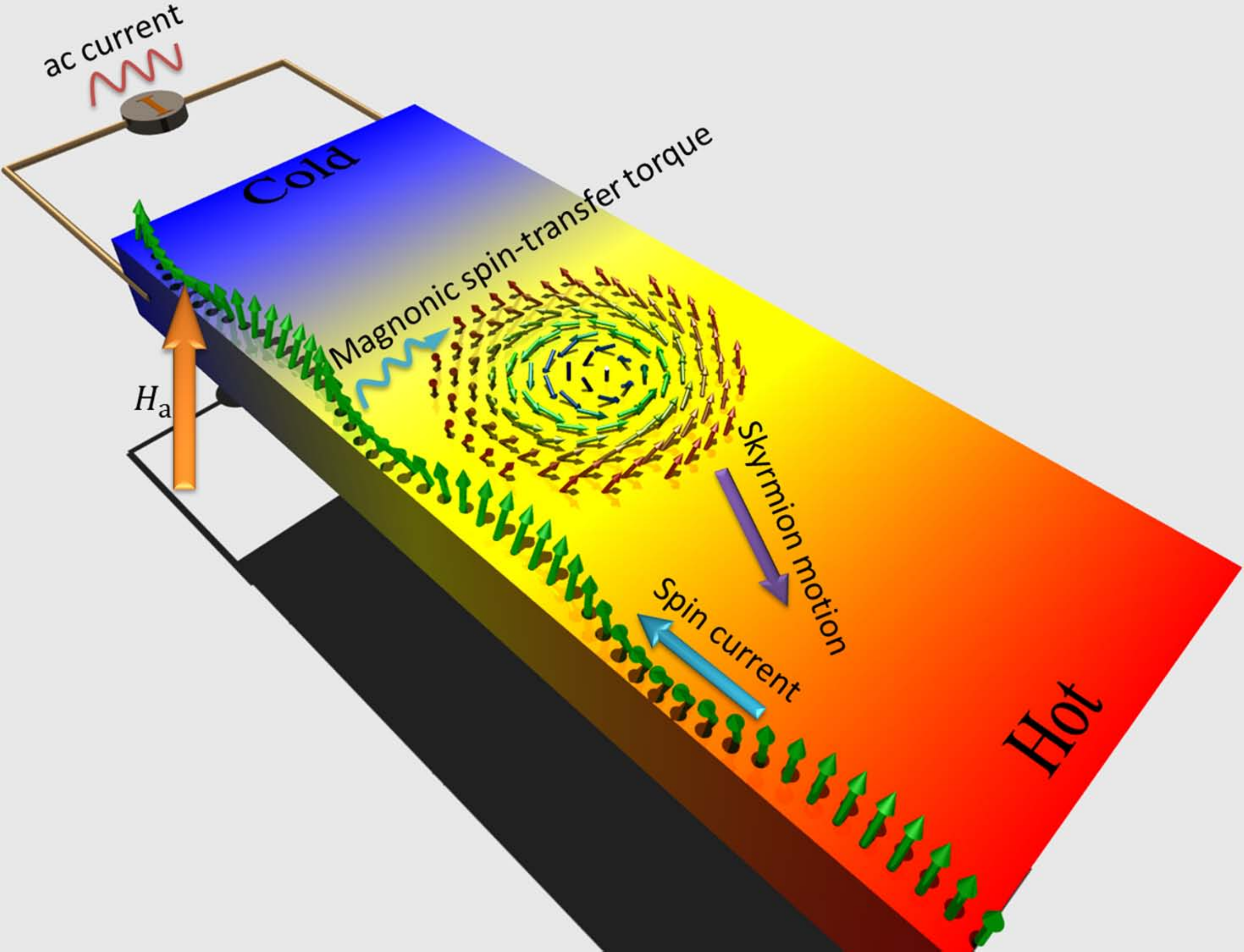,width=\columnwidth}
\caption{(color online) Schematic view of the motion of a skyrmion in the presence of a temperature gradient. The temperature gradient excites a magnon current flowing from the hot to the cold region, which drives the skyrmion opposite to the magnon current direction. The skyrmion motion changes the electric polarization, which generates an ac current when the insulating magnet is embedded in a closed circuit.
} \label{f1}
\end{figure}

We study the dynamics of skyrmions in insulating chiral magnets under  the presence of a thermal gradient and find that skyrmions move from the cold to the hot region due to the magnonic spin transfer torque generated by the temperature gradient (spin Seebeck effect). We also present a phenomenological description of skyrmion dynamics with a monochromatic magnon current. We then compute the dynamics of electric polarization ${\bf P}$ that produces an electric ac current density $d{\bf P}/dt$ when the insulator is embedded in a closed circuit. The role of the skyrmion motion on the ac current generation is clarified. Finally, when the temperature in the hot region is high enough to induce a local paramagnetic state, we find that skyrmions are continuously generated   by thermal fluctuations and then diffuse into the cold region.

We consider a thin film of insulating chiral magnet with thickness $d$ (Fig. 1), which is described by the Hamiltonian \cite{Bogdanov89,Bogdanov94,Rosler2006,Han10,Rossler2011},
\begin{equation}\label{eq1}
\mathcal{H}=\int d\mathbf{r}^2 \left[\frac{J_{\rm{ex}}}{2}(\nabla \mathbf{n})^2+D\mathbf{n}\cdot(\nabla\times \mathbf{n})-\mathbf{H}_a\cdot\mathbf{n} \right],
\end{equation}
where $J_{\rm{ex}}$ is the exchange interaction, $D$ is the Dzyaloshinskii-Moriya (DM) interaction \cite{Dzyaloshinsky1958,Moriya60,Moriya60b}, which is generally present in magnets without inversion symmetry and $\mathbf{n}$ is a unit vector denoting the spin direction and $\mathbf{r}=(x,\ y)$. The system is assumed to be uniform along the $z$ direction for thin films. The external field $\mathbf{H}_a=H_a\hat{z}$ is perpendicular to the film and its magnitude is fixed at $H_a=0.6D^2/J_{\mathrm{ex}}$ to stabilize the skyrmion phase. \cite{Rossler2011} The spin dynamics is governed by the Landau-Lifshitz-Gilbert equation \cite{Tatara2008}
\begin{equation}\label{eq2}
{\partial _t}{\bf{n}} =- \gamma {\bf{n}} \times ({{\bf{H}}_{\rm{eff}}+\tilde{\mathbf{H}}}) + \alpha {\partial _t}{\bf{n}}\times \mathbf{n}, 
\end{equation}
where $\mathbf{H}_{\rm{eff}}\equiv-\delta \mathcal{H}/\delta {\bf{n}}={J_{\mathrm{ex}}}{\nabla ^2}{\mathbf{n}} - 2D\nabla  \times {\mathbf{n}} + {\mathbf{H}}_a$, $\gamma=a^3/(\hbar s)$, $a$ is the lattice constant and $s$ the magnitude of local spins. $\tilde{\mathbf{H}}$ is the fluctuating magnetic field that introduces thermal fluctuations and consequently satisfies the local fluctuation-dissipation theorem: $\langle \tilde{\mathbf{H}} \rangle =0$ and
\begin{equation}\label{eq3}
\langle \tilde{H}_{\mu}(\mathbf{r}, t)\tilde{H}_{\nu}(\mathbf{r}', t') \rangle=\frac{2 k_B T(\mathbf{r})\alpha}{d \gamma} \delta_{\mu,\nu}\delta(\mathbf{r}'-\mathbf{r})\delta(t'-t),
\end{equation}
where $T(\mathbf{r})$ is the local temperature at $\mathbf{r}$, $k_B$ is the Boltzmann constant and $\mu, \nu=x,\ y,\ z$. Here we have assumed a local equilibrium for magnons characterized by spatially dependent temperature $T(x)$ and that the Seebeck effect is driven by magnons. \cite{Xiao2010,Adachi11} Depending on the kinetics of the phonons and magnons, and on their coupling, other effects may come into play such as the phonon drag spin Seebeck effect \cite{Adachi10} and the nonlinear temperature profile along the sample. \cite{Agrawal13,Tikhonov2013}

 \begin{figure}[t]
\psfig{figure=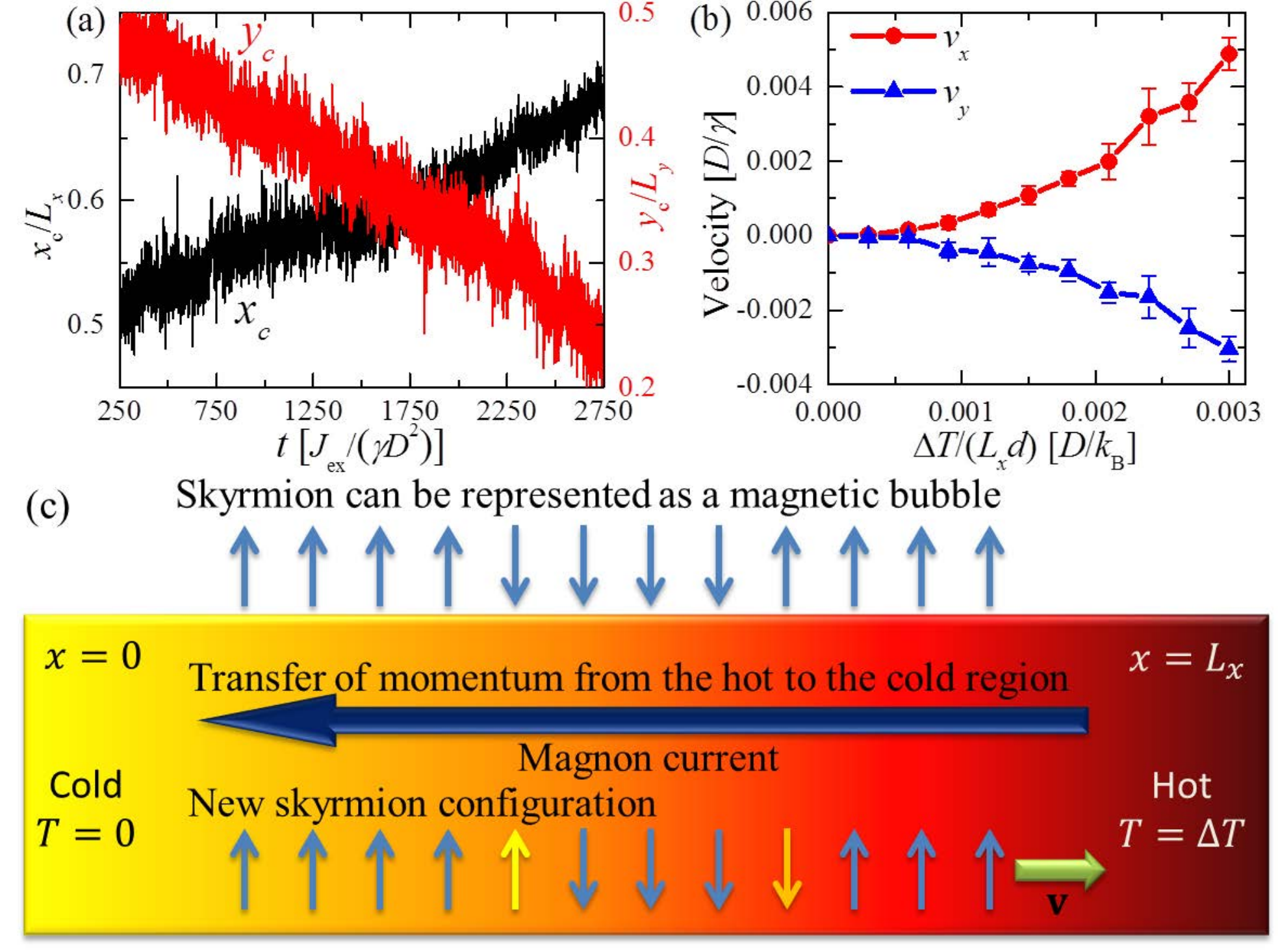,width=\columnwidth}
\caption{(color online) (a) Trajectory of the skyrmion center of mass in the presence of a temperature gradient. Here $\Delta T/(L_x d)=0.003$ and $\alpha=0.1$. (b) Average drift velocity as a function of temperature gradient. The curves are obtained by averaging over $8$ independent runs. (c) Schematic view of the magnonic spin transfer torque and the dynamics of the skyrmion.
} \label{f2}
\end{figure}

We first study the skyrmion dynamics  in the presence of a  finite temperature gradient: $T(x=0)=0$ and $T(x=L_x)=\Delta T$.  Note that $\Delta T$ is not high enough to induce a local paramagnetic state on the  hot side, and to excite new skyrmions thermally. First we prepare a stationary skyrmion in the ferromagnetic background state at $T(x)=0$, and then turn on the temperature gradient. We calculate numerically \cite{szlin13skyrmion1,noteNumerics} skyrmion position
$\mathbf{r}_c={\int d \mathbf{r}^2 [{\bf{n}}\cdot({\partial _x}{\bf{n}} \times {\partial _y}{\bf{n}})]\mathbf{r}}/{\int d \mathbf{r}^2 [{\bf{n}}\cdot({\partial _x}{\bf{n}} \times {\partial _y}{\bf{n}})]}$,
and its  velocity $\mathbf{v}=\dot{\mathbf{r}}_c$. The corresponding trajectory  is shown in Fig. \ref{f2}(a).  Surprisingly,  the skyrmion moves from the cold  towards the hot region. The average velocity depends linearly on the temperature gradient as shown in Fig. \ref{f2}(b). Similar behavior is obtained for many skyrmions. \cite{supplement}

In magnetically ordered insulators, the heat current is carried both by  magnons and phonons \cite{Meier2003,Yan2012}. In the same way a finite temperature gradient induces an electron current in metals and produces the conventional Seebeck effect,
it also produces a magnon current in magnetic insulators (magnon flow) that leads to the so-called spin Seebeck effect.  This effect has been measured in metals \cite{Uchida2010a}, semiconductors \cite{Jaworski2010} and insulators \cite{Uchida2010b}. The magnon current carries a magnetic moment and interacts with the skyrmion via the magnonic spin transfer torque. Since a skyrmion is topologically equivalent to a magnetic bubble,  we represent the skyrmion as a bubble domain of downward spins in a spin up background to illustrate the basic mechanism [see Fig.  \ref{f2} (c)].  When the magnon passes through the skyrmion, the magnetic bubble is displaced by one lattice constant in the opposite direction of the magnon current to conserve the magnetic moment, which explains the results in Fig. \ref{f2} (a). Note that the net entropy still flows from the hot to the cold region, because the density of magnons is higher than the density of skyrmions.

\begin{figure}[t]
\psfig{figure=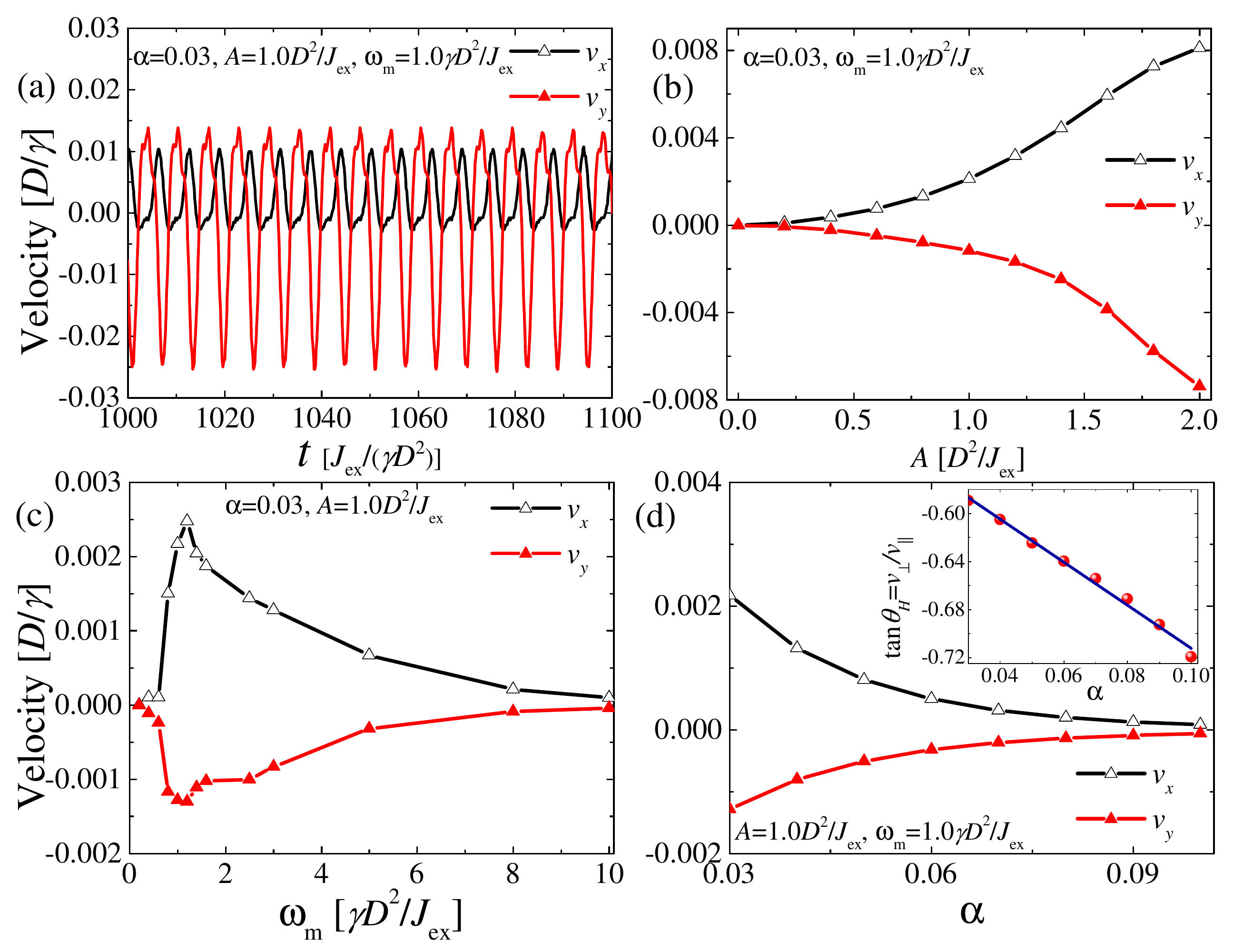,width=\columnwidth}
\caption{(color online) (a) Oscillation of the skyrmion velocity with a nonzero dc component in the presence of a magnon current (b-d). Dependence of the skyrmion velocity on the amplitude $A$ and frequency $\omega_m$ of the ac magnetic field applied at the right edge of the sample $x=L_x$, which acts as a source of magnon current, and on the Gilbert damping $\alpha$. The inset in (d) is the Hall angle as a function of $\alpha$. The line is a fit to $\tan\theta_H$ in the main text.
} \label{f3}
\end{figure}

The thermally excited magnon has all frequency components. We will start by considering the interaction between a monochromatic magnon current and a skyrmion in the absence of noise $\tilde{H}=0$. \footnote {In simulations, the magnon current is created by applying an ac magnetic field along the $x$ direction $H_{x,\ ac}(x=L_x)=A\sin(\omega_{m}t)$ on the right edge of the sample. The field can only excite magnons if $\omega_m$ is larger that the magnon gap: $\omega_m> \gamma H_a/(1+\alpha^2)$.} The magnon decay produced by damping and the decay length can be obtained from the magnon dispersion. The initial distance between the skyrmion and the magnon source  is $L_x/2$ to ensure that the magnon current is not fully damped out before interacting with the skyrmion: $L_x<4\sqrt{J_{\mathrm{ex}}\gamma(\omega_m-H_a\gamma)}/(\omega_m\alpha)\approx 1\ \mathrm{\mu m}$ for typical parameters, where $\omega_m$ is the magnon frequency. The skyrmion oscillates with frequency $\omega_m$ and drifts in the  opposite  direction to the magnon current [Fig. \ref{f3} (a)].  The drift velocity increases with the amplitude, $A$, of the ac driving field [see Fig.~\ref{f3} (b)], and it has a nonmonotonic dependence on $\omega_m$. It reaches its maximum value when the wavelength of the magnon, $2\pi/k_m$ with $k_m$ given by $\omega_m=\Omega({\bf k}_m)$, is of the order of the skyrmion size [Fig. \ref{f3} (c)]. This effect could be used to measure the skyrmion size. The magnon current is quickly damped for increasing Gilbert damping $\alpha$  and the drift velocity decreases [Fig. \ref{f3} (d)]. For  thermally excited magnons, the oscillation of the skyrmion velocity smears out while the drift motion remains.

To understand our numerical results we introduce  a phenomenological description of the interaction between skyrmions and a magnon current. For this purpose we separate $\mathbf{n}$ into a slow component, $\mathbf{n}_s$, corresponding to the skyrmion motion  and a fast component, $\tilde{\mathbf{n}}$, induced by the magnon current: $\mathbf{n}=\mathbf{n}_s+\tilde{\mathbf{n}}$. By following the procedure in Ref. \cite{Kovalev2012} and coarse-graining over $\tilde{\mathbf{n}}$, Eq. \eqref{eq2} becomes
\begin{equation}\label{eq6}
{\partial _t}{{\bf{n}}_s} =  \alpha {\bf{n}}_s \times {\partial _t}{\bf{n}}_s - \gamma {{\bf{n}}_s} \times (J{\nabla ^2}{{\bf{n}}_s} - 2D\nabla  \times {{\bf{n}}_s} + {\bf{H}}_a)-{\partial _\mu }\mathbf{J}_\mu+\mathbf{\Gamma},
\end{equation}  
with $\mathbf{\Gamma} \propto \gamma D k_m |\tilde{\bf{n}}|^2 $ being the contribution from the DM interaction. Here the tensor $\mathbf{J}_\mu=\gamma J_{\mathrm{ex}}\tilde{\mathbf{n}}\times\partial_\mu\tilde{\mathbf{n}}$ is the spin current density (note that $\mu=x,y$ is the spatial coordinate). The total spin is not conserved, ${\partial _\mu }\mathbf{J}_\mu\neq 0$, because of  the presence of DM interaction and external magnetic fields. $\mathbf{\Gamma}$ can be neglected for  typical parameters, $D\ll J_{\mathrm{ex}}/a$,  and magnon wavelength much shorter than the skyrmion size [$|\Gamma|/|{\partial _\mu }\mathbf{J}_\mu| \approx  D/(J_{\mathrm{ex}} k_m)\ll 1$]. We note that $|{\bf n}_s|=1$ and ${\bf n}_s \times {\tilde {\bf n}}=0$ in the linear approximation. In addition, ${\bf n}_s \times \partial_{\mu} {\tilde {\bf n}} \simeq 0$ because this term is linear in ${\tilde n}$ implying that it must vanish after coarse-graining in time over the fast oscillations of ${\tilde n}$. Therefore,  we can rewrite the magnon current as $\mathbf{J}_\mu=J_\mu \mathbf{n}_s$ with $J_\mu=\gamma J_{\mathrm{ex}}(\tilde{\mathbf{n}}\times\partial_\mu\tilde{\mathbf{n}})\cdot\mathbf{n}_s$, and the magnon spin transfer torque is  given by $ {\partial _\mu }\mathbf{J}_\mu\approx J_\mu\partial_\mu \mathbf{n}_s$.~\cite{Kovalev2012}

Equation \eqref{eq6} is equivalent to the case of skyrmion motion driven by a spin polarized current $- J_\mu$ in metals. The magnonic spin transfer torque is always adiabatic because the magnetic moment of magnons is antiparallel to the local moment $\mathbf{n}_s$. The $\Gamma$ term cannot be neglected when the magnon wavelength becomes comparable to the skyrmion size. In this case, skyrmions get  distorted by the interaction with magnons. This effect can be accounted for by adding a non-adiabatic spin transfer torque term to Eq.~\eqref{eq6}
\begin{equation}\label{eq8}
{\partial _t}{\bf{n}}_s =  - \gamma {\bf{n}}_s \times {{\bf{H}}_{\mathrm{eff}}} + \alpha {\bf{n}}_s \times {\partial _t}{\bf{n}}_s + ({\bf{v}}_m\cdot\nabla ){\bf{n}}_s - {\beta }{\bf{n}}_s \times ({\bf{v}}_m\cdot\nabla ){\bf{n}}_s.
\end{equation}
The equation for drift motion is derived from Eq.~\eqref{eq8}~\cite{szlin13skyrmion2,Iwasaki2013} by treating the skyrmion as a rigid particle~\cite{Thiele72}: 
\begin{equation}\label{eq9}
-\hat z \times ( \mathbf{v}_m + \mathbf{v}) + \eta (\alpha \mathbf{v}-\beta \mathbf{v}_m ) = 0,
\end{equation}
where $\eta=\eta_{\mu}={\int {{d}}{\mathbf{r}^2}{({\partial _{\mu}}{\bf{n}}_s)^2}}/(4\pi)$ is the form factor of a skyrmion. The effective magnon velocity $\mathbf{v}_m$ and the parameter $\beta$ can be estimated by fitting the skyrmion trajectory obtained from simulations to Eq. \eqref{eq9}. Note that the skyrmion velocity has a component transverse to the temperature gradient as displayed in Figs. \ref{f1} and \ref{f2} (a). To linear order in $\alpha\ll 1$, we obtain a Hall angle   $\tan{\theta _H} \equiv v_{\perp}/{v_{\parallel}} =- \beta \eta  - ( \eta  + {\beta ^2}{\eta ^3})\alpha$ for the skyrmion motion, where $v_{\perp}$ and $v_{\parallel}$ are the velocity components  perpendicular and parallel to the flow direction of the magnon current. The inset of Fig.~\ref{f3}~(d) shows the dependence of $\theta_H$ on $\alpha$. By fitting this curve, we obtain $\beta\approx 0.4$ and $\eta\approx 1.4$, which indicates that the non-adiabatic spin transfer torque plays an important role in the interaction between the magnon current and the skyrmion when $2 \pi/ k_m$ is comparable to the skyrmion size.

 \begin{figure}[t]
\psfig{figure=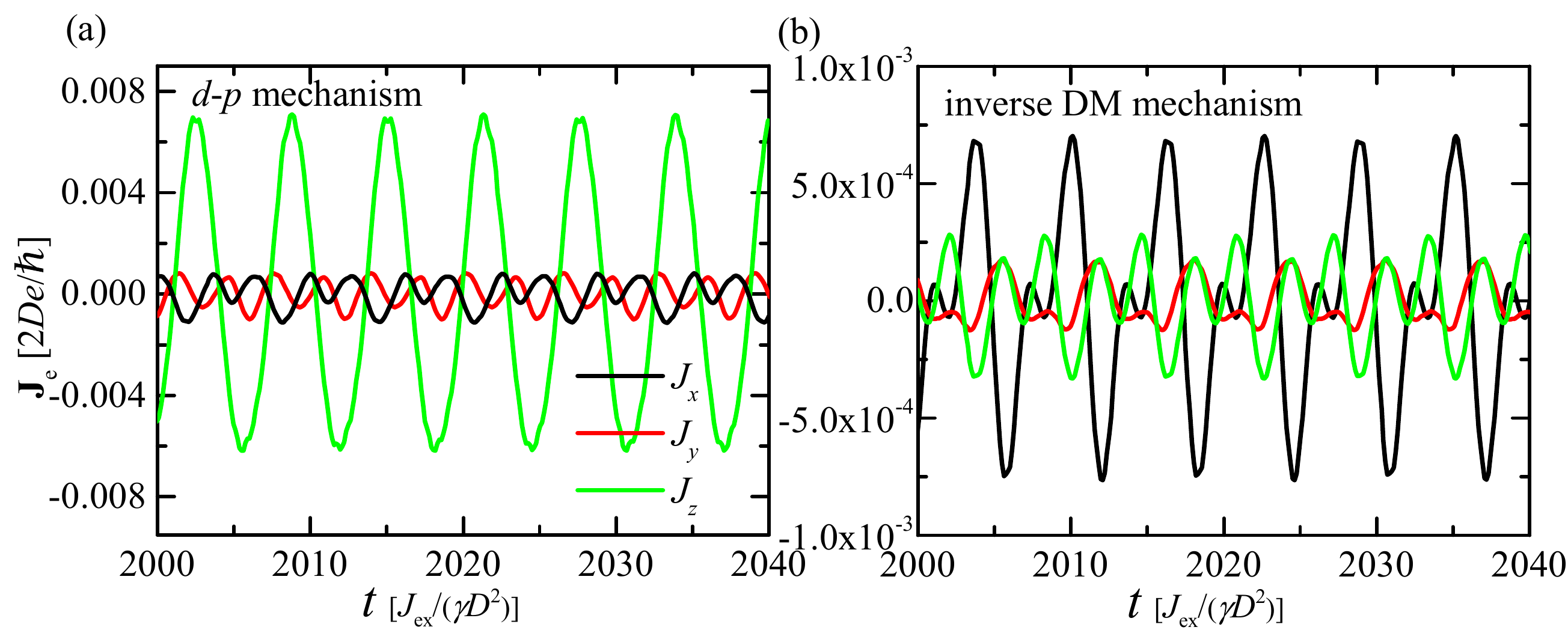,width=\columnwidth}
\caption{(color online) ac current induced by the skyrmion motion and magnon current, according to (a) the $d$-$p$ hybridization and (b) the inverse DM mechanism. Here $\alpha=0.03$, $A=1.0 D^2/J_{\mathrm{ex}}$, $\omega_m=1.0\gamma D^2/J_{\mathrm{ex}}$,
} \label{f4}
\end{figure}

We next study the electric polarization induced by the skyrmion motion. For $\mathrm{Cu_2OSeO_3}$, it was demonstrated experimentally that the mechanism for the generation of electric polarization is $d$-$p$ hybridization~\cite{Seki2012,Seki2012b}, which arises from the interaction between a ligand (oxygen) ion and a transition metal (copper) ion with a single magnetic moment.~\cite{Jia2006,Jia2007,Yang12} Because the electric polarization depends on the direction of the external magnetic field, we introduce a new coordinate frame with the $z$-axis parallel to $\mathbf{H}_a$ and the $x$-axis along the $[\bar{1}01]$ direction. We consider the case $\mathbf{H}_a \parallel [110]$. The electric polarization is given by ${\bf{P}}_{dp} = \frac{P_0}{2} ( - 2 {n_x}{n_y},n_z^2- n_x^2, 2{n_y}{n_z} )$,
where $P_0\approx 50\ \mathrm{\mu C/m^2}$ for $\mathrm{Cu_2OSeO_3}$ \cite{Seki2012b}. In other potential realization of skyrmion lattices, electric polarization could also be induced by the mechanism known as inverse DM effect \cite{Katsura2005,Cheong2007}: $\mathbf{P}_{\mathrm{IDM}}=\mathcal{P}_0[ {\hat e_x } \times ({\bf{n}} \times {\partial _x }{\bf{n}})+{\hat e_y } \times ({\bf{n}} \times {\partial _y }{\bf{n}})]$.
If the sample is embedded in a circuit, the time dependent polarization leads to an ac electric current density $\mathbf{J}_e=\partial_t \mathbf{P}$. For the inverse DM mechanism, this current density can be expressed in terms of the magnon current $\mathbf{J}_{\mu}$, 
$\mathbf{J}_{e,\mathrm{IDM}}(\omega) =  \mathrm{Im}\left[ \frac{{\mathcal{P}_0\omega }}{\gamma }[{\hat e_x } \times {{\bf{J}}_x }(\omega )+{\hat e_y } \times {{\bf{J}}_y }(\omega )]\right]$.
The ac electric current induced by a monochromatic magnon current is shown in Fig.~\ref{f4}. The electric current has the frequency of the magnon current. The time average of the induced current vanishes and the amplitude of the ac electric current density is about $5\times 10^3\ \mathrm{A/m^2}$ for typical parameters. For the $d$-$p$ hybridization mechanism, $\mathbf{P}_{dp}$ is finite for collinear spin textures and an ac electric current is still induced in the absence of skyrmions. The  presence of skyrmions produces a small change in the electric current. In contrast, the inverse DM mechanism is absent   for collinear spin textures ($\mathbf{P}_{\mathrm{IDM}}=0$) and  skyrmions are then necessary to induce electric polarization. Because the skyrmion  contribution to spin fluctuations is much smaller than that from magnons, the magnitude of the ac current for the inverse DM mechanism is much smaller than that for the $d$-$p$ hybridization mechanism [see  Fig.~\ref{f4}].

Finally, we increase the temperature in the hot region by $\Delta T$ such that the hot region is in the paramagnetic phase, where skyrmions are created and destroyed dynamically by thermal fluctuations. Because of the presence of a temperature gradient, some of the created skyrmions diffuse into the cold region and stabilize there. Skyrmions are driven towards the cold region by the newly created skyrmions in the hot region because of an effective repulsion between them. Meanwhile, the magnon current drives the skyrmions from the cold to the hot region. However, the repulsive interaction dominates and skyrmions keep diffusing from the hot to the cold region, as observed in our simulations. \cite{supplement}

Skyrmions  can also be driven by a temperature gradient in metallic magnets. In this case, the skyrmion motion   induces an emergent electric field given by $\mathbf{E}={\hbar }\mathbf{n}\cdot \left(\nabla \mathbf{n}\times\partial_t\mathbf{n} \right)/{2 e}$, \cite{Zang11} i.e. there is a skyrmion Hall voltage induced by the spin Seebeck effect. The conventional Seebeck effect for electrons also produces a longitudinal and a Hall voltage in the presence of an external magnetic field. This electronic contribution is dominant because the electron carrier density  is much higher than the density of skyrmions. 

In inhomogeneous systems, there is a random pinning potential for skyrmions. Then, a threshold temperature gradient or a magnon current density is required to move the skyrmions. We can estimate the threshold temperature gradient by using the depinning current density measured for metallic magnets, expecting that the pinning potential  is similar for insulators and metals. By using typical parameters and the results shown in Fig.~\ref{f2}, we estimate the velocity to be $0.1\ \mathrm{m/s}$ for a temperature gradient $\Delta T/L_x\approx 0.04$ K/nm and a film thickness of $d\approx 10$ nm. The required current density to achieve a similar velocity in metallic magnets by a spin-polarized current  is $J\approx 10^9\ \mathrm{A/m^2}$, i.e.,  much larger than the typical depinning current $J\approx 10^6\ \mathrm{A/m^2}$. Thus, the skyrmion can be depinned at a temperature gradient larger than $\Delta T/L_x\gtrsim 4\times 10^{-5}$ K/nm. We note that local thermal fluctuations are also helpful for skyrmions to creep from the pinning sites.   

To summarize, we have studied the generation of an ac current and the motion of skyrmions in insulating chiral magnets subject to a temperature gradient. The skyrmions move from the cold to the hot region because of the magnonic spin transfer torque. When the temperature of the hot region is high enough to induce a local paramagnetic state, skyrmions are created by thermal fluctuations and diffuse into the cold region.  The generation of the ac current by a thermal gradient does not depend on the presence of skyrmions for the magneto-electric coupling that arises from the $d$-$p$ hybridization mechanism. However, for the inverse Dzyaloshinskii-Moriya mechanism, an ac current is induced only when skyrmions are present to render the spin texture non-collinear. Our results indicate that skyrmions in insulating chiral magnets are promising for spin caloritronics applications.

\noindent \textit{Acknowledgments --} Computer resources for numerical calculations were supported by the Institutional Computing Program at LANL. This work was carried out under the auspices of the NNSA of the US DOE at LANL under Contract No. DE-AC52-06NA25396, and was supported by the US Department of Energy, Office of Basic Energy Sciences, Division of Materials Sciences and Engineering. 

\noindent \textit{Note added in proof --} After completion of the present work, we become aware of the similar results on the motion of a Skyrmion in the presence of a temperature gradient in Ref. \cite{Kong2013}. However the usual diffusion of Skyrmions and the induced electric current due to the magnetoelectric coupling are not discussed in Ref. \cite{Kong2013}. The non-adiabatic magnonic spin transfer torque proposed in Eq. \eqref{eq8} is derived systematically in Ref. \cite{Kovalev2014}.

%\bibliography{reference}

\end{document}